\input harvmac
\overfullrule=0pt

\Title{\vbox{
\baselineskip 10pt
\rightline{hep-th/9808016}
\rightline{NSF-ITP-98-082}
\rightline{EFI-98-30}
\rightline{RU-98-xx}}}
{\vbox{\centerline{AdS Dynamics from Conformal Field Theory}}}

\bigskip

\baselineskip=12pt
\centerline{%
Tom Banks$^{*\,\dagger}$%
\footnote{$^a$}{\tt  banks@physics.rutgers.edu},
Michael R. Douglas$^{*\,\dagger\,\&}$%
\footnote{$^b$}{\tt  mrd@physics.rutgers.edu},
Gary T. Horowitz$^{\star\,\dagger}$%
\footnote{$^c$}{\tt  gary@cosmic.physics.ucsb.edu},
and 
Emil Martinec$^{\#\,\dagger}$%
\footnote{$^d$}{\tt  e-martinec@uchicago.edu}
}
\medskip
\centerline{$^\dagger$ Institute for Theoretical Physics }
\centerline{Santa Barbara, CA   93106-4030}
\medskip
\centerline{$^*$ Department of Physics and Astronomy}
\centerline{Rutgers University, Piscataway, NJ 08855}
\medskip
\centerline{$^\star$ Department of Physics}
\centerline{University of California, Santa Barbara, CA 93106-4030}
\medskip
\centerline{$^\#$ Enrico Fermi Institute and Department of Physics}
\centerline{University of Chicago}
\centerline{5640 S. Ellis Ave., Chicago, IL 60637-1433}
\medskip
\centerline{$^\&$ I.H.E.S., Le Bois-Marie, Bures-sur-Yvette, 91440 France}

\baselineskip=16pt
 
\vskip 1cm
\noindent

We explore the extent to which a local string theory dynamics 
in anti-de Sitter space can be determined from its proposed 
Conformal Field Theory (CFT) description.
Free fields in the bulk are constructed from
the CFT operators, but difficulties are encountered when
one attempts to incorporate interactions.  We also discuss
general features of black hole dynamics as seen
{}from the CFT perspective. In particular, we 
argue that the singularity of $AdS_3$ black holes is resolved
in the CFT description.

\Date{8/98}

%
\def\dalam{\mathop{\mathpalette\MAKEdalamSIGN\nabla}\nolimits}
\def\MAKEdalamSIGN#1#2{%
\setbox0=\hbox{$\mathsurround 0pt #1{#2}$}
\dimen0=\ht0 \advance\dimen0 -0.8pt
\hbox{\vrule\vbox to\ht0{\hrule width\dimen0 \vfil\hrule}\vrule}}

\def\AdSS5{$AdS_5$}
\def\AdS5s5{AdS_5 \times S^5}
\def\gy{g_{\sst\rm YM}}

\def\G(#1){\Gamma(#1)}
\def\alphaprime{\alpha'}

\def\C|#1{{\cal #1}     }
\def\(#1#2){(\zeta_#1\cdot\zeta_#2)}


\def\xxx#1 {{hep-th/#1}}

\def\npb#1(#2)#3 { Nucl. Phys. {\bf B#1} (#2) #3 }
\def\rep#1(#2)#3 { Phys. Rept.{\bf #1} (#2) #3 }
\def\plb#1(#2)#3{Phys. Lett. {\bf #1B} (#2) #3}
\def\prl#1(#2)#3{Phys. Rev. Lett.{\bf #1} (#2) #3}
\def\physrev#1(#2)#3{Phys. Rev. {\bf D#1} (#2) #3}
\def\ap#1(#2)#3{Ann. Phys. {\bf #1} (#2) #3}
\def\rmp#1(#2)#3{Rev. Mod. Phys. {\bf #1} (#2) #3}
\def\cmp#1(#2)#3{Comm. Math. Phys. {\bf #1} (#2) #3}
\def\mpl#1(#2)#3{Mod. Phys. Lett. {\bf #1} (#2) #3}
\def\ijmp#1(#2)#3{Int. J. Mod. Phys. {\bf A#1} (#2) #3}

\parindent 25pt
\overfullrule=0pt
\tolerance=10000

%
%
\def\journal#1&#2(#3){\unskip, \sl #1\ \bf #2 \rm(19#3) }
\def\andjournal#1&#2(#3){\sl #1~\bf #2 \rm (19#3) }

\def\ie{{\it i.e.}}

\def\cf{{\it c.f.}}
\def\etal{{\it et.al.}}

\def\sst{\scriptscriptstyle}

\def\frac#1#2{{#1\over#2}}

\def\vev#1{\langle#1\rangle}
\def\d{\partial}

\def\inbar{\,\vrule height1.5ex width.4pt depth0pt}
\def\IC{\relax\hbox{$\inbar\kern-.3em{\rm C}$}}
\def\IR{\relax{\rm I\kern-.18em R}}
\def\IP{\relax{\rm I\kern-.18em P}}

\def\nth{$n^{\rm th}$}
%
%

\def\npb#1#2#3{Nucl. Phys. {\bf B#1} (#2) #3}

\def\plb#1#2#3{Phys. Lett. {\bf #1B} (#2) #3}
\def\prl#1#2#3{Phys. Rev. Lett. {\bf #1} (#2) #3}
\def\physrev#1#2#3{Phys. Rev. {\bf D#1} (#2) #3}
\def\prd#1#2#3{Phys. Rev. {\bf D#1} (#2) #3}

\def\rmp#1#2#3{Rev. Mod. Phys. {\bf #1} (#2) #3}
\def\cmp#1#2#3{Comm. Math. Phys. {\bf #1} (#2) #3}
\def\cqg#1#2#3{Class. Quant. Grav. {\bf #1} (#2) #3}
\def\mpl#1#2#3{Mod. Phys. Lett. {\bf #1} (#2) #3}
\def\ijmp#1#2#3{Int. J. Mod. Phys. {\bf A#1} (#2) #3}

\catcode`\@=11
\def\slash#1{\mathord{\mathpalette\c@ncel{#1}}}
\overfullrule=0pt

\def\GG{{\cal G}}

\def\underrel#1\over#2{\mathrel{\mathop{\kern\z@#1}\limits_{#2}}}

\catcode`\@=12


%

\def\vev#1{\left\langle #1 \right\rangle}

\def\tr{{\rm tr}}

\def\exp{{\rm exp}}

\def\bh{{\sst BH}}

\def\lpl{\ell_{{\rm pl}}}
\def\lplthree{\ell_{{\rm pl}}^{{\sst (3)}}}
\def\lstr{\ell_{s}}

\def\gstr{g_{ s}}
\def\adsthree{{${\it AdS}_3$}}
\def\rads{R}
\def\adsschw{{\sst\rm AdS-Schw}}
\def\corr{{\sst\rm corr}}
\def\hag{{\sst\rm Hag}}
\def\btz{{\sst\rm BTZ}}

\def\lz{{L_0}}
\def\lzb{{{\bar L}_0}}

\def\R {{\bf R}}

\def\pp{\varphi}

\def\IH{\relax{\rm I\kern-.18em H}}

%
\nref\malda{J. Maldacena, hep-th/9711200.}
\nref\spectrum{S. Ferrara, C. Fronsdal, and A. Zaffaroni, hep-th/9802203;
O. Aharony, Y. Oz, Z. Yin, hep-th/9803051; J. de Boer, hep-th/9806104.}
\nref\inter{S. Lee, S. Minwalla, M. Rangamani and N. Seiberg,
hep-th/9806074;
E. D'Hoker, D. Z. Freedman and W. Skiba, hep-th/9807098.  }
\nref\bhent{See, e.g., I. Klebanov and A. Tseytlin, hep-th/9604089,
\npb{475}{1966}{164}.}
\nref\bdhmii{T. Banks, M. Douglas, G. Horowitz, 
and E. Martinec, to appear.}
\nref\stu{L. Susskind, L. Thorlacius, and J. Uglum, hep-th/9306069,
\prd{48}{1993}{3743}; Y. Kiem, H. Verlinde, and E. Verlinde, hep-th/9502074,
\prd{52}{1995}{7053}.}
\nref\btzbh{M. Banados, C. Teitelboim, and J. Zanelli, \prl{69}{1992}{1849}.}
\nref\ho{G. Horowitz and H. Ooguri, hep-th/9802116,
\prl{80}{1998}{4116}.} 
\nref\ml{M. L\"uscher and G. Mack, \cmp {41} {1975} {203}.} 
\nref\adswave{M. G\"unaydin and N. Marcus, \cqg{2}{1985}{L11};
H. Kim, L. Romans, and P. van Nieuwenhuizen, \prd{32}{1985}{389}.}
\nref\ftb{S.
 Weinberg, {\it The Quantum Theory of Fields : Foundations},
Cambridge Univ. Press (1995).}
\nref\gkp{S.S. Gubser, I.R. Klebanov, and A.M. Polyakov, hep-th/9802109.}
\nref\wittenads{E. Witten, hep-th/9802150; hep-th/9803131.}%
\nref\calc{D.Z. Freedman, S. Mathur, A. Matusis, and L. Rastelli,
hep-th/9804058.} 
\nref\wbs{N. N. Bogoliubov and D. V. Shirkov, {\it Introduction to the theory 
of quantized fields}, John Wiley (1980).}
\nref\nishijima{K. Nishijima, {\it Fields and Particles: field theory and
dispersion relations}, W. A. Benjamin (1974).}
\nref\susswit{L. Susskind and E. Witten, hep-th/9805114.}
\nref\uvir{A. Sen, hep-th/9605150, \npb{475}{1996}{562};
T. Banks, M.R. Douglas, N. Seiberg, hep-th/9605199, \plb{387}{1996}{278};
N. Seiberg, hep-th/9606017, \plb{384}{1996}{81}.}
\nref\fms{D. Friedan, E. Martinec, and S. Shenker, \npb{271}{1986}{91}.}
\nref\mbg{M. Bianchi, M.B. Green, S. Kovacs, and G. Rossi, hep-th/9807033.}
\nref\emilstr{E. Martinec, talk at Strings'98;
http://www.itp.ucsb.edu/online/strings98/martinec.}
\nref\glf{R. Gregory and R. Laflamme, hep-th/9301052,
\prl{70}{1993}{2837}.}
\nref\hopo{G. Horowitz and J. Polchinski, hep-th/9612146, \prd{55}{1997}{6189}.}
\nref\bkl{V. Balasubramanian, P. Kraus, A. Lawrence, and S. Trivedi, to appear.}
\nref\juan{J. Maldacena, hep-th/9803002, \prl{80}{1998}{4859}.}
\nref\dt{M. R. Douglas and W. Taylor IV, hep-th/9807225.}
\nref\maldatalk{J. Maldacena, talk at Strings'98;\hfil\break
http://www.itp.ucsb.edu/online/strings98/maldacena.}
\nref\berk{M. Berkooz, hep-th/9807230.}
%

\newsec{\bf Introduction}

It has recently been proposed \malda\
that string/M theory on $AdS_{d} \times K$
(where $K$ is a suitable compact space) is equivalent to a
conformal field theory (CFT)
`living on the boundary'%
\foot{Although we adhere to current usage, the CFT does not really
live on the boundary; see section 4.1.}
of the anti-de Sitter space.
Some evidence for this conjecture
comes from the agreement of the
spectrum of supergravity fluctuations 
with the spectrum of operators in the conformal
field theory \spectrum.  In part this follows simply from the isomorphism
between the symmetry groups of the two theories, but the correspondence
also correctly matches the multiplicities of irreducible representations.
The extension of the conjecture to all
of string/M theory is based on our expectation that this is the unique
quantum completion of supergravity.
Further evidence comes from the agreement of certain perturbative
interactions \inter, and the  ability of the CFT to explain the entropy of AdS
black holes \bhent.

Given this AdS/CFT correspondence, it is natural to ask to what extent
local spacetime physics on $AdS_{d}$
can be recovered from the CFT, 
which is a lower-dimensional field theory.
Below we discuss several aspects of this question
(a number of details are postponed to a future publication \bdhmii). 
First we show that
one can construct free quantum fields corresponding to all of the modes
of string theory in  this background. These are operators in the CFT which
depend on position in the AdS spacetime, and satisfy the usual causality
conditions. One might have worried that since the CFT operators are causal
with respect to the boundary causal structure which does not include the
AdS radial coordinate, it would be difficult to construct operators which
commute whenever they are spacelike separated in AdS. We will see that there
is an essentially unique way of avoiding this difficulty
in the large
$N$ limit in which the string theory becomes free. In this limit, the
combination of large $N$ factorization and group theory determines the
operator algebra of the CFT to be that of creation and annihilation
operators of free string modes on $AdS_d \times K$.  From these we can
construct local free fields in a unique way. These operators
turn out not to be local when interactions are included. It is not yet clear
whether our expressions for the spacetime operators can be modified to 
remain local. Nor have we understood the precise nature of the nonlocality
and the extent to which it becomes invisible at low energies.
In making these constructions,
we work mostly with $AdS_5\times S^5$, but similar
arguments should work  for backgrounds of the form $AdS_3\times S^3 \times M$.

Our construction of local fields provides a clue for the discussion of
local motions in spacetime.  In particular, the symmetry group of the CFT
generates the action of the corresponding spacetime symmetries by vector
fields acting on our local operators.  We use the intuition derived from
this correspondence to discuss black hole dynamics. We explain how certain objects which 
fall into an AdS black hole can be described in terms of the CFT. The
black hole itself is described by a high energy state which looks approximately
thermal. The object falling in is described by a localized excitation in the
gauge theory. As it evolves in CFT time, its scale size grows.
We show that the time for this excitation to 
expand to the size of the typical thermal wavelength (and hence become
indistinguishable from the background) agrees with the time  in AdS for
the object to fall from a large radius to the vicinity of the horizon.
The natural
time evolution in the CFT corresponds to evolution with respect to the
external Schwarzschild time of the black hole. So one never sees the
object cross the horizon. We argue that observers that cross the horizon
are described by evolving the state with respect to an operator different
{}from the CFT Hamiltonian. This prescription is a precise realization
of the idea of {\it black hole complementarity} \stu.
The CFT Hilbert space does not
break up into a tensor product of spaces inside and outside of the
horizon.  Rather there is a single Hilbert space describing both
inside and outside.  Physics as seen by different observers corresponds
to acting on this space with different classes of operators.  The operators
corresponding to an external observer do not commute with those of
an infalling observer.
In the case of the three dimensional BTZ black hole \btzbh,
one can use the local symmetries to identify a suitable operator. Since this
operator is another conformal generator which acts unitarily on the CFT
Hilbert space, the evolution is still nonsingular. We are thus led to the 
conclusion that in the context of the AdS/CFT correspondence, quantum
effects indeed resolve the BTZ singularity!  The singularity seen in the
classical supergravity description of physics as seen by this observer, must
be an artifact of the large $N$ limit.


\newsec{\bf Linearized supergravity fields in the large N limit}

\subsec{ A Little Scaling Argument}

We will be working, in this and the following section,
with the $AdS_5 \times S^5$ system, in the regime where there
is a clear separation between the long distance expansion and
the perturbation expansion (as noted in the introduction, similar
arguments should work for backgrounds 
of the form $AdS_3 \times S^3 \times M$).
The dual CFT is the ${\cal N} = 4$ Super Yang Mills theory (SYM).  
We will be studying the 't Hooft limit of the SYM
and the $1/N$ expansion around it, though we would like the 't Hooft
coupling to be large since the radius $R$ of the AdS is given by
${R^4 \over \alphaprime^2} = \gy^2 N \equiv \lambda $.
Free Type IIB string theory on $\AdS5s5$ should then be the leading term
in the planar expansion of ${\cal N} = 4, \ d =4$ SYM theory.
In order to do perturbative string theory on a space of low curvature
we take $\lambda$ large but independent of $N$ and expand amplitudes
in inverse powers of $N$.  
That is, we are in the regime $1\ll \gy^2 N \ll N$.
The perturbative gauge theory gives
a natural classification of properties of operators in the large $N$ limit
according to the number of powers of the trace which they involve.

Let $\{ {\cal O}_i \}$ be a complete basis\foot{
We assume that the equations of motion of the gauge theory are used to
eliminate redundant operators.}
 of single trace operators.  Using
a standard normalization, their full and connected Green's functions
satisfy the following scaling relations for even $n$:
\eqn\scala{ \vev{{\cal O}_{i_1} \ldots {\cal O}_{i_n} } \sim N^n}
\eqn\scalb{ \vev{{\cal O}_{i_1} \ldots {\cal O}_{i_n} }_c \sim N^2}
Note that in CFT the VEV of any nonunit operator vanishes.  As a consequence
the full and connected three point functions are the same and
both scale like $N^2$.  Full $2k+1$ point functions scale like
$N^k$ and connected ones like $N^2$.

If we define rescaled operators by $O_i = {1\over N}{\cal O}_i$, then
the rescaled operators have unit normalized two point functions and the
$1/N$ expansion of their connected Green's functions looks (combinatorially) 
like
a perturbation expansion around a free field theory, in which
the $O_i$ are independent free fields.  Note also that
in the operator product expansion $O_i O_j = \sum {1\over N} C_{ij}^k O_k$,
we should expect the coefficients $C_{ij}^k$ to be of order one for
large $N$.

Multiple
trace operators, defined as $O_{i_1 \ldots i_n} = {1\over N^n} {\cal O}_{i_1} 
\ldots {\cal O}_{i_n}$\foot{There are actually many multiple trace 
operators for any given set of single trace operators, corresponding
to all the primary fields in the complete operator product expansion.}
will have two point functions normalized to one.   Connected Green's functions
of products of single trace operators with appropriate multiple trace
operators will be of order one in the large $N$ limit.  In terms of
the analogy with perturbation theory described in the previous paragraph,
the multiple trace operators behave like composites of the 
``free fields''.  The operator product of two single trace operators
contains terms of order one with multiple trace operators.

\subsec{A Little Group Theory}

To leading order in the expansion
in $\gstr$ at fixed $\alpha'$, the conventional description of
string theory on a given background suggests that it can be viewed as
an infinite number of free quantum field theories propagating on this
background.  Our first goal is to construct this `free string
field theory' from the quantum variables of the CFT.
Before doing so, we must remark on the issue of gauge fixing, which
we will not address in this paper.

In the standard field theoretic approach to constructing the Hilbert
space of quantum gravity, one must choose a gauge, and in most
gauges, introduce ghosts and a BRST operator.  
We restrict attention
to a class of gauges which are defined by covariant
({\it e.g.} De Donder) conditions
on the components of the supergraviton fields, in AdS space.
In the present paper we will consider only the Green's functions of
fields which are scalars in AdS, to leading order in perturbation theory.
In the above class of gauges, the gauge fixing and ghost Lagrangians will
not involve the scalar degrees of freedom.   Thus, if we concentrate on
scalars, and their leading tree level self-interaction we should be
able to ignore the issue of gauge fixing.  
The details\foot{in which however, the devil often resides.  
One interesting problem which we will not
address is the relation between the diffeomorphism invariance
of supergravity and the gauge invariance of the SYM theory.  Since most of
the (super) Killing diffeomorphisms of the model are accompanied by gauge 
transformations in their action
on the non gauge invariant part of the SYM Hilbert space, we suspect 
that a close relation does in fact exist.} of leading order BRST
quantization of the full system should be straightforward but tedious.

The Hilbert space of the free string theory is the Fock space
constructed from a collection
of unitary irreducible representations of the $AdS_5 \times S^5$
super-isometry group 
$SU(2,2|4)$.  These single particle representations
all have positive {\it energy}, in terms of the generator of the standard
global time translations in $AdS$ space.   It was pointed out in \ho\
that the corresponding Hilbert space for the CFT is obtained
by quantization of the gauge theory on $S^3 \times \R$,
for it is only here that the conformal group is implemented in
a unitary fashion.%
\foot{The quantization of the gauge theory on $\R^{3,1}$
carries a realization of the Lie algebra of conformal transformations,
but not a unitary implementation of the group.}
As shown in \ml,
the generator $H \equiv K^0 + P^0$ of the conformal group is positive
in the class of unitary representations of the conformal group obtained
by quantizing a unitary CFT on $S^3 \times \R$;
thus we map $H$ to the global time translation generator
on the $AdS$ side.
We note also that the authors of \ml\ have argued that in a generic
CFT the global time coordinate $t$ cannot be taken periodic.  Thus as
in \ho\ we assume that we are working on the universal cover of $AdS$ space.

A scalar particle in $AdS_5 \times S^5$ can be labelled by an $S^5$ 
spherical harmonic which we call $Y$, an ``orbital'' angular
momentum $J$ on $S^3$, and a frequency $\omega > 0$ .  In terms of these,
the mass of the particle is determined by the equation for the Casimir
operator of $SU(2,2|4)$ \adswave.
The particle states are in one-to-one correspondence
with the positive frequency solutions of the scalar wave equation on
$AdS_5 \times S^5$.   In (dimensionless) coordinates where
the $AdS_5$ metric is
\eqn\metric{
  ds^2 = - (1 + r^2) dt^2 + {dr^2 \over (1+r^2)} + r^2 d\Omega_3^2\ ,
}
this equation reads
\eqn\kg{
  \left[- {1 \over (1+r^2)} \partial_t^2  + 
	{1\over r^3}\partial_r [r^3 (1 + r^2) \partial_r] + 
	{1\over r^2} L_3^2 +L_5^2\right]\,\psi=0 \ .
}
Here, $L_3^2$ (respectively, $L_5^2$) is the square 
of the angular momentum operator on $S^3$ ($S^5$).
Of course we must also pick the solution which is normalizable in the
usual Klein-Gordon norm on surfaces of constant $t$.

At the risk of being pedantic, we remind the reader of a straightforward
consequence of the correspondence between particle states and wavefunctions.
The spacetime symmetry group action on supergravity fields is implemented
by Killing vector fields ${\cal L}_a$.  The solutions $\psi_{\omega ,J,Y}$
of the scalar wave equation satisfy
\eqn\wvfcns{
  {\cal L}_a \psi_{\omega ,J,Y } = 
	D_{\omega ,J,Y }^{\rho ,K ,Z}    
	\psi_{\rho ,K ,Z}\quad ,
}
where $D$ is the same matrix which implements the operation of the Hilbert
space operators on the states:
\eqn\stts{
  L_a \vert \omega ,J,Y \rangle = D_{\omega ,J,Y }^{\rho , K , Z }   
	\vert \rho , K , Z \rangle\ .
}       

In the CFT description, states with the transformation properties of single
scalar supergraviton states are constructed by acting with conformal
primary operators on the conformally invariant vacuum.  Note that because
of the positivity of $H$, operators which satisfy $[H, O] = \omega O$ with
negative $\omega$ must annihilate the vacuum.  Thus, if we Fourier expand
a local operator 
\eqn\posfreq{
  O(\Omega_3 , t) = \sum_{n = 0}^{\infty}  [O_{\omega_n} (\Omega_3 )
	e^{ - i \omega_n t} + h.c. ]\ , 
}
then $O_{\omega}$ annihilates the vacuum and the states are created by 
$O_{\omega}^{\dagger}$ .  Note that a field in a given representation
of the conformal group on $S^3 \times \R$ actually contains only a discrete
set of frequencies $\omega_n$.

By construction, the transformation properties of the operators 
$O_{\omega}^{\dagger}$ under the Killing symmetries of the 
background spacetime  are the same
as those of supergraviton creation operators.
We would now like to argue that at leading order in $1/N$ they obey the same 
algebra.  Indeed, we have argued that the single trace operators, $O_i$
have only connected two point functions in the large $N$ limit.  Furthermore,
the positive frequency components of these fields annihilate the vacuum.
As a consequence, we can extract the commutators of the positive and negative
frequency components from the norm of the states created from the vacuum
by the negative frequency operators.  
These norms are determined by group theory
and are therefore the same for supergravity and the CFT.

It is now an easy generalization of the arguments of Weinberg \ftb\ 
that, to leading order in $1/N$,
there is a unique set of local fields in $\AdS5s5 $ 
which can be constructed from the single trace conformal fields of
the SYM theory.  These fields satisfy the fundamental equation
\eqn\fundeq{{\cal L}_a \phi = i [L_a , \phi ]\ ,}
which says that the symmetry operators of the CFT act on them like
the appropriate Killing vectors of the spacetime.  

The fields have the expansion
\eqn\expansion{
  \phi (t, \Omega_3 , \Omega_5 ,r) = \sum_{{\omega > 0}, J, Y} [
	\psi_{\omega ,J,Y} (r) e^{- i\omega t} Y_J (\Omega_3 ) 
	Y_Y (\Omega_5 ) O_{\omega , J, Y} + h.c. ]\ .
}
where the $Y$ index on $O$ denotes the conformal primary operator  associated
with the $S^5$ spherical harmonic $Y$.
Below we will have occasion to use a condensed notation for this formula.
If we rewrite this in terms of the local fields $O$ it has the
generic appearance
\eqn\intform{\phi (x) = \int_b G(x,b) O(b)}
where $x$ is a point in $\AdS5s5 $ and $b$ a point on its boundary.
The Green's function $G(x,b)$ is implicitly defined by $\expansion$.
It is a solution of the homogeneous Klein-Gordon equation and its
properties will be further explored in \bdhmii.

The two-point function of $\phi(x)$ is by construction that
of the supergravity fields, since the wavefunctions in \expansion\
solve the wave equation (determined by group theory),
and the algebra of the modes of the CFT operators $O$ is 
just that of creation and annihilation operators at
this order in $1/N$.  In particular, the propagation is 
causal.  The extent to which this can be maintained at higher
orders in $1/N$ will be discussed below.


\newsec{\bf The Effect of Interactions}

We have thus, modulo the presumably technical questions of gauge fixing,
constructed the free local string field theory of this string theory
compactification as a set of operators in the CFT which is supposed
to encode the exact dynamics of the theory.
We now want to generalize these considerations to the interacting
theory.  Before beginning to calculate, we wish to point out a general
property of the CFT which may be crucial for understanding why this
theory is different from local field theory.

The algebra of the single trace operators is only approximately that
of independent creation and annihilation operators.   In the full theory 
it is highly constrained and not at all free.
It is easy to see that the representation space
for this algebra should be much smaller than the algebra
of independent creation and annihilation operators,
due to the operator product relations 
$O_i O_j = \sum {1\over N} C_{ij}^k O_k$.%
\foot{An amusing example with some similar properties 
is an affine Lie algebra at level $k$,
where the large $k$ limit corresponds to our large $N$ limit.
At leading order in $1/k$, the currents are abelian and their
modes are independent creation and annihilation operators.
At next-to-leading order, there are relations among them
given by the affine Lie algebra structure constants;
this is why $c<{\rm dim}\,G$.  Note also that the interaction
Lagrangian is not locally expressed in terms of the currents,
although here there is a simple way out, by passing to the
group field.  Below we will argue that there is no such
simple fix for the string field theory.}
This is, we believe,
an indication that the theory has many fewer degrees of freedom than
one expects from a field theory.   Note that we say field theory
rather than String Field Theory.  This is because there are operator
product relations even between operators in short representations
of the superconformal algebra.  Thus not even all of the would-be
supergraviton creation and annihilation operators are independent
in the CFT.   

With these philosophical comments dispensed with, we are nearly
ready to begin discussing interactions.  However we must first see whether
our benign neglect of the problem of gauge fixing can affect the
discussion.  We believe that it will not (within the classes of gauges
we have discussed above) if we restrict our attention to three point functions
of AdS scalar fields at lowest order in perturbation theory.  
The only graphs which contribute to these Green's functions at this order
involve scalar propagators, and the triple scalar coupling in the Lagrangian.
Since none of these objects are changed by a change of gauge within the
allowed class, we can compute them without a full discussion of gauge
fixing.

The formula \expansion\ for the scalar fields 
in lowest order perturbation theory
already implies a contribution to the connected three point function
in leading order in $1/N$.  This comes precisely from the $1/N$ contribution
to operator products of single trace operators.
Thus, the fields defined by \expansion\ have a connected 3 point Wightman 
function
\eqn\wthree{
  \vev{\phi_1 (x_1 ) \phi_2 (x_2 ) \phi_3 (x_3 )}_c = \int G(x_1 , b_1 )
  G(x_2 , b_2 ) G(x_3 , b_3 ) \vev{O_1 (b_1 ) O_2 (b_2 ) O_3 (b_3 ) }_c\ .}
There are two interesting questions to ask of this formula.  Does it
coincide with the lowest order perturbative formula of supergravity?  Does
it reproduce the boundary correlation functions of supergravity discussed
in \refs{\gkp,\wittenads}?  
Of course, a positive answer to the first of these
questions would obviate the need to ask the second.

It is easy to see however that the answer to the first question is no.
If we apply the appropriate scalar wave operator to any of the three
legs of this Green function, it vanishes.  In supergravity, the fields
satisfy nonlinear wave equations of the schematic form
\eqn\waveq{\dalam \phi_1 = \phi_2 \phi_3\ ,}
where we have taken only the relevant trilinear scalar coupling into account.
In evaluating the Green's functions perturbatively, we find 
a nonvanishing connected three point function only by taking into
account this leading nonlinearity in the field equations.  Thus,
the leading connected three point function satisfies
\eqn\nlwveq{\dalam_1 \vev{\phi_1(x_1)\; \phi_2(x_2)\; \phi_3(x_3)\; } = 
  \vev{ \phi_2 \phi_3 (x_1)\; \phi_2(x_2)\; \phi_3(x_3)\; }\ ,}
which is inconsistent with our formula.   We will return below to
the question of whether it is possible to modify the field
to obtain the supergravity formula for the Green functions.

First however, we turn to the second question.  
Gubser \etal\ \gkp\ and Witten \wittenads\
have presented a connection between supergravity 
Green functions and CFT correlators.
There is a graphical version of this prescription which resembles the
LSZ formula for S-matrix elements in Minkowski space.  The CFT
correlators are extracted from supergravity by calculating the graphs
for Green functions at points in the bulk $\AdS5s5 $ space and then
replacing the external lines by a special homogeneous Green function
$G_W$ of the scalar wave operator which implements delta function boundary
conditions on the $S^3 \times \R$ boundary.    

At large spacelike distance, the Feynman Green function $G_F (x,y)$
for a scalar of mass $m$ behaves like $r^{ - d_m}$ where 
$d_m = 2+(4+m^2)^{1/2}$.
Here we have used the coordinates \metric\ and go to infinity by taking
$r$ large with other coordinates fixed.  The function $G_W (x,b)$ is
only defined when one of its arguments is on the boundary.  We can
obtain it from $G_F$ by the formula
\eqn\limit{G_W (x,b) = \lim_{y \rightarrow b} G_F (x,y) r_y^{d_m}\ .}
Thus, we can obtain the GKP/W correlators by multiplying the 
Feynman Green functions by powers of $r$ and taking a limit in
which the external points go to the boundary.
\eqn\lsz{
  G_{GKP/W} (b_1 \ldots b_n ) = 
	\lim_{y_i \rightarrow b_i} 
		\prod r_{y_i}^{d_{m_i}} G(y_1 \ldots y_n)\ .
}
There may be some subtleties in this limiting procedure because the
interior points of the diagram are integrated over all of AdS space
and the measure of integration is concentrated on the boundary.
Indeed, in the calculations of \calc,
great care had to be taken
to obtain correct results.  We have not studied the question of
whether these subtleties invalidate the simple formula \lsz.
Note by the way that although we have described the derivation of
this formula perturbatively, it would appear to make sense 
(again modulo questions about gauge variance) 
as a nonperturbative relation.  The formula is also, despite its
apparent dependence on a particular coordinate system, coordinate
invariant.  That is, we can easily replace the explicit coordinate
factors in the equation by factors of the geodesic distance between
the points $y_i$ and some arbitrarily chosen interior point.  Up
to an overall constant factor for each external leg (a sort of
wave function renormalization), the answer does not depend on the
choice of interior point.

Let us now apply this procedure to the time ordered Green's functions
of the fields $\phi $ defined in \expansion.  Since our limiting procedure
involves the variation of only the spacelike variable $r$, we can apply it
to each term in the time ordered product.   Thus, we might as well
study the limit of Wightman functions with a particular ordering of
the operators.   If we harmonically expand each field as in \expansion,
we see that the harmonically transformed Wightman function 
is just the ordered product of factors of the form $\psi_{\omega , J, Y}
 (r) O_{\omega , J , Y}$.  Now we exploit the fact that for fixed $Y$
(and fixed values of all other implicit labels on the operator), 
the large $r$ behavior of all of the $\psi$ functions is the same as 
that of the Feynman Green function with the same value of the mass.
Thus, up to an overall constant,
(which we can absorb in the definition of our LSZ formula) the large
$r$ behavior of the Wightman function will be the Wightman function in
the CFT of the conformal fields $O_Y (b)$ (we are implicitly assuming
that the limit commutes with the integrals which define the harmonic
expansion).  In view of the remarks at the beginning of this paragraph,
the same will be true of the time ordered functions.

To summarize, what we have proved, without recourse to any expansion,
is that the Green functions of the field defined by \expansion\ reproduce
all of the GKP/W correlation functions.  This is rather more than
we bargained for -- a field satisfying the linearized field equations
computes the exact nonperturbative ``S matrix''.

In fact, this result shows us that the GKP/W correlation functions
cannot be thought of as an S-matrix in the sense of an overlap between
exact multiparticle eigenstates of the CFT Hamiltonian.  If such an
interpretation were possible, then the Green functions of $\phi (x)$
with one variable in the bulk and the rest taken to the boundary:
\eqn\formfac{
  G_{GKP/W} (x, b_1 , \ldots b_n ) =  
	lim_{y_i \rightarrow b_i} 
		\prod r_{y_i}^{d_{m_i}} G(x, y_1 \ldots y_n)
}  
would have to be interpreted as the form factor or matrix element%
\foot{The ambiguity in splitting the boundary points into two groups
in this expression is another indication that we are not calculating
an S-matrix.  The spacelike limit we are taking does not naturally
separate asymptotic points into future and past.}
\eqn\formfacb{\vev{b_1 , \ldots b_k \vert \phi (x) \vert b_{k+1} \ldots b_n}.}
The free field equation for $\phi$ would then imply that this expression
could be nonzero only if the two states differed by addition of a
single particle of the mass carried by the field.  
Then, using the formula \lsz\ we would conclude that the amplitude 
$\vev{O(b_1 ) \ldots O(b_k) O(b_{n+1} ) O(b_{k+1}) \ldots O(b_n )}$
obeyed the same restrictions.  This is absurd, because of the
complete symmetry of these amplitudes in the boundary points, unless
the amplitudes vanish for $n > 2$.  But of course we know that this
is not the case.   

We emphasize that this conclusion is tied to the use of the global
time of AdS space as evolution parameter.  We have, for example, used
global time ordering to define our amplitudes.  
As mentioned above, the spacetime
geometry of this problem suggests that no sensible S-matrix 
interpretation of the asymptotic limit of the global time evolution is to be
expected.  Geodesics simply do not separate from each other asymptotically
in time.  D-brane black hole physics suggests that these amplitudes
do have a sensible S-matrix interpretation in terms of the Minkowski
time evolution generated by $P^0$ in the conformal field theory
(as opposed to the global evolution defined by $P^0+K^0$).
Motions along the flows of the corresponding $AdS$ 
generator do not correspond to geodesic motion in $AdS$ space
(rather they are like the flows of Rindler time).
However they do appear to correspond to motions of incoming and outgoing 
particles in the asymptotically flat geometry of which $\AdS5s5 $ is
the near horizon limit.

We now turn to the question of whether we can modify our definition
of the field to order $1/N$ in order to make it coincide with
the fields of perturbative supergravity, 
at least in some low energy approximation.  
This is the same as asking whether the field can be made local.
Indeed, textbook arguments \wbs\
imply that local perturbations of free field theory can always be 
understood as perturbations of the Hamiltonian by integrals of 
a local density.  Correspondingly, the connected $n$ point functions
of such perturbed local fields are directly connected to the nonlinear
terms in the equations of motion which they satisfy.   

This implies that the field we have constructed in \expansion\ is not
local at next-to-leading order in $1/N$.  It satisfies a free field
equation but has a connected 3 point function at order $1/N$.%
\foot{Since our field is not local, its time ordered Green's functions 
will not be covariant under symmetries of AdS space
which change the definition of the global time.  We emphasize
that the {\it physical} argument for covariance of these Green's
functions requires us to contemplate perturbations of
the system by a local external source.  Such perturbations are not
allowed in a quantum theory of gravity because they destroy
the covariant conservation law for the stress tensor.  This
is perhaps the most primitive reason for believing that the
quantum theory of gravity is holographic.}

It is of course well known that, to leading order in derivatives, the
only consistent local perturbation of the free supergravity fields is
that of interacting supergravity.   More generally, any higher derivative
covariant correction to the supergravity action would be an acceptable 
perturbation at leading order.  Schematically, any local field
with a connected 3 point function would have to satisfy an
equation of the form
\eqn\ife{\dalam \phi = {1\over N} \phi^2\ .}
If we call the field of \expansion\ 
$\phi_0$ and write $\phi = \phi_0 + \Delta$,
then we can solve for $\Delta$ to leading order in $1/N$:
\eqn\soldelt{\Delta = {1\over N}(\int G_R (x,y) \phi_0^2 (y) + \phi_1 )}
where $G_R$ is the retarded inhomogeneous Green function for the scalar wave operator
and $\phi_1$ is a solution of the free wave equation.

If we keep only the first term in $\Delta$ and (correctly to this order) treat
$\phi_0$ as a local free field, then we are solving the usual Yang-Feldman equation
\nishijima\ and we reproduce the expected local supergravity Wightman
function.  The other terms of order $1/N$ come from the connected part of the
$\phi_0$ three point function in CFT, and from insertions of $\phi_1$, with $\phi_0$ 
treated as a free field.  Thus, in order to remain with just the supergravity
formula for the three point function we must have a cancellation
\eqn\ldthree{
  \vev{\phi_0 (x_1 ) \phi_0 (x_2 ) \phi_0 (x_3 )}_c + {1\over N}
  (\vev{\phi_1 (x_1 ) \phi_0 (x_2 ) \phi_0 (x_3 )}_c 
+ permutations ) = 0}
Here, permutations, refers to the two other possible positions for $\phi_1$
in the Green's function.
Remember that the connected three point function of $\phi_0$ is of order $1/N$
so that these terms are of the same order.  Note also that to this
order, the expectations of products of $\phi_0$ in the last three terms
are to be evaluated as if these were free fields in $AdS$ space.

Because this equation involves integrals over the entire boundary of
the singular CFT correlators, making sense of it requires additional
information, such as a prescription for analytic continuation.
Such issues will be discussed in \bdhmii\ .  At present we are unable to tell
whether a field $\phi_1$ satisfying this equation can be constructed.
The issue of the existence of local fields including interactions
remains to be clarified.

However, 
an indication that modifications along these lines are on the
wrong track comes from the expression for the $SU(2,2|4)$
generators that one would deduce from the 
free fields \expansion.  Each of the infinite tower of
free fields $\phi_\alpha$ ($\alpha$ labels all the 
quantum numbers besides those of $\AdS5s5$)
has its own generator of $SU(2,2|4)$; the full generator
that implements the isometries on the supergravity fields
(or even the string field of closed string field theory)
has the form
\eqn\wronggen{
  \GG_a^{{\sst \rm SUGRA}}=\sum_\alpha \phi^\dagger_\alpha
	i[L_a,\phi_\alpha]\ .
}
This expression, which contains terms of arbitrarily high
order in the SYM fields, has essentially no relation
to the proper generators in terms of the CFT (super)stress tensor 
$T_{\sst \rm CFT}$ and superconformal Killing fields $v_a$
\eqn\rightgen{
  \GG_a^{{\sst \rm CFT}}=\int_b v_a\cdot T_{{\sst \rm CFT}}
}
which are bilinear in the SYM field strengths.  Small $1/N$ corrections
to the relation between the $\phi_\alpha$ and the
SYM fields will not repair this disparity; the SYM
theory has a fundamentally different structure,
with far fewer degrees of freedom, from which
the supergravity fields are built as highly composite objects.

In addition, the fact that the free fields we have
constructed compute the exact interacting correlation functions
(once the correct CFT operator algebra is used) in such a simple
way, suggests to us that the operator solutions of the interacting
supergravity field equations in the CFT Hilbert space are needlessly
complicated objects.

Nevertheless,
an important reason to attempt a construction of local fields 
is the following.
To the extent that our framework precisely reproduces supergravity
it will be exactly local, and such a construction must exist.
On the other hand,
previous work on the black hole information paradox strongly suggests
that the full theory must contain some breakdown of locality.
Therefore, the attempt to construct
local field theory should break down at some level, and it would
be extremely interesting to know where this happens.

The limit in which locality is recovered
is likely to be highly context-dependent. 
There cannot be a universal limit in which high-frequency
excitations are averaged out.  For instance, an examination
of well-separated objects in M-theory would lead one to
conclude that membrane and fivebrane excitations should
be integrated out below the Planck scale.  Yet it is precisely
these objects (when bound together in combination with
supergravitons) that are responsible for generating the extremely
small gap -- which can be arbitrarily smaller than the Planck scale --
observed in near-extremal black holes in
four and five dimensions.  The recovery of approximately
local physics in the interacting theory is likely to
be a rather delicate issue.

\newsec{Aspects of black hole dynamics}

While a detailed understanding of the interacting theory requires
further investigation, qualitative aspects of  black hole dynamics
can already be understood.
This section is divided into five parts.  We begin by setting the stage for
our discussion of black holes: We  consider first the relation
between scale in the CFT and radial position in $AdS$, and then
some general features of the spectrum of string theory on $AdS_d\times S^p$.
The next two parts discuss objects falling toward a black hole, 
including  wrapped fundamental strings and waves on a three brane.
Finally, we explain how one might describe an observer falling into a
black hole in terms of the CFT.

\subsec{IR-UV duality}

There is an important correspondence between small/large radial position
in AdS and IR/UV phenomena in the CFT \refs{\malda,\susswit}.%
\foot{Such a correspondence was foreshadowed by the observed IR/UV 
duality of brane probes of background geometry \uvir,
and indeed by similar well-known phenomena in perturbative
string theory (\cf\ \fms).}  One can gain insight into this by considering
the frequently asked question: ``Where are the branes?''
According to the map \intform,
the superconformally invariant vacuum of the CFT maps
to the vacuum state in AdS, suggesting that the branes 
are {\it everywhere}.  A qualitative way to see that the 
branes fill the entire AdS spacetime is to consider
the operator $\tr[X^2]$ in the example of $\AdS5s5$.
This operator measures the mean square radial position
of the branes.  In the strong coupling limit, it
has a large anomalous dimension 
$\Delta\sim (\gstr N)^{1/4}\sim \rads/\lstr$.
Fluctuations in the radial position are diagnosed by
the correlator
\eqn\radfluct{
  \vev{\tr[X^2(z)]\;\tr[X^2(0)]}\sim |z|^{-2\Delta}\ .
}
Thus, locality in the $3+1$ dimensions of the gauge theory is in
direct conflict with locality in the radial position.
As we consider shorter and shorter distance scales in the gauge theory,
fluctuations in the radial position become arbitrarily large.
Conversely, fluctuations on the longest length scales probe the 
`center' of the AdS space\foot{Once we have chosen a given generator
$P^0 + K^0$ as
the CFT Hamiltonian, the `center' of AdS can be defined as the point where the
corresponding Killing vector has minimal norm.}.
We see qualitatively why a scale in the CFT corresponds
to a radial location $r$
in the bulk. We also see why $\tr[X^2]$ acquires
a large anomalous dimension -- it is the
radial coordinate on $AdS_5$.
Operator insertions
on the branes are pointlike, hence extreme UV in character;
this is why they are conventionally regarded as acting
on the boundary of AdS.
On the other hand, a given field configuration, such as an instanton, has a scale,
putting it at the corresponding radial position \mbg.
As a consequence, the CFT does not really `live' on the boundary
of the AdS spacetime; rather, it fills the bulk.
One can regard it as a {\it representation}
of M-theory dynamics, much as worldsheet CFT is
a representation of perturbative string dynamics.
Indeed, the perturbative string also fills spacetime
due to its quantum fluctuations; pointlike UV perturbations --
the vertex operators -- represent perturbations at the 
asymptotic boundary of spacetime, yet one would
not say that the string worldsheet resides at the (conformal)
boundary of Minkowski space.

\subsec{The supergravity spectrum}

We now consider some general features of
the spectrum of string theory on $AdS_d\times S^p$.%
\foot{We thank J. Maldacena for helpful discussions
which improved our understanding of this spectrum.} 
At sufficiently high energy, the typical state in the gauge theory 
describes an $AdS_d$ Schwarzschild black hole (which is
constant on $S^p$), with horizon size $r_+>\rads$
and positive specific heat \wittenads.
For lower energies, such that $\lstr<r_+<\rads$,
one has a phase of ordinary $(d+p)$-dimensional Schwarzschild
black holes \emilstr; the black hole localizes
on $S^p$ due to the Gregory-Laflamme instability \glf.
This latter phase is stable microcanonically,
as is the `Hagedorn' phase that appears below
the (correspondence point) energy $E_\corr$ where $r_+\sim\lstr$ \hopo.
These two phases would be missed in an analysis of the
canonical ensemble where, because
of the negative specific heat, once the energy reaches
the string scale the external heat bath pumps energy
into the system until it reaches the threshold to form
an AdS Schwarzschild black hole.
At very low energies, one expects a gas of supergravity 
particles in AdS to prevail.

We illustrate with two examples.
For $\AdS5s5$, the hierarchy of scales is ($\lpl$ here
denotes the 10d Planck scale)
\eqn\hifive{\eqalign{
  E_\adsschw&\sim{\rads^7}{\lpl^{-8}}
	\sim N^2\rads^{-1}\cr
  E_\corr&\sim{\lstr^7}{\lpl^{-8}}
	\sim N^2\rads^{-1}(\gstr N)^{-7/4}\cr
  E_\hag&\sim\lstr^{-1}(\gstr N)^{9/4}\sim \rads^{-1}(\gstr N)^{5/2}\ ,
}}
such that the entropy is 
$S(E)\sim N^2(\rads E/N^2)^{3/4}$ for $E>E_\adsschw$,
and one has five-dimensional AdS Schwarzschild black holes;
$S(E)\sim N^2(\rads E/N^2)^{8/7}$ for $E_\corr<E<E_\adsschw$,
and one has ten-dimensional Schwarzschild black holes;
$S(E)\sim(\lstr E)\sim\rads E(\gstr N)^{-1/4}$
for $E_\hag<E<E_\corr$, where fundamental strings dominate the entropy; and
$S(E)\sim (RE)^{9/10}$ for $E<E_\hag$, where the entropy is dominated by a
gas of supergravity particles in $AdS_5 \times S^5$.
We have assumed strong coupling $\gstr N>1$,
otherwise the supergravity gas and the 10d black hole 
phase disappear.

Similarly, in $AdS_3\times S^3\times M$, there is a corresponding
set of scales (here $\lplthree$ denotes the 3d Planck scale, $R$ is the $AdS_3$
radius, $\gstr$ is the 6d string coupling, and $k=Q_1Q_5$)
\eqn\hithree{\eqalign{
  E_\btz&\sim(\lplthree)^{-1}\sim {k}{\rads^{-1}}\cr
  E_\corr&\sim k\lstr^{-1}(\gstr^2 k)^{-3/4}\sim 
	{ k}{\rads^{-1}}(\gstr^2k)^{-1/2}\cr
  E_\hag&\sim\lstr^{-1}(\gstr^2 k)^{5/4}\sim
	\rads^{-1}(\gstr^2k)^{3/2}\ .
}}
For completeness, we note that the Kaluza-Klein modes and
string winding modes on $M$ have typical scales
$\rads^{-1}(\gstr Q_5)^{1/2}$ and $\rads^{-1}(\gstr Q_1)^{1/2}$,
respectively.

One might ask what regime of parameters is described
by the $S^k(M)$ symmetric orbifold CFT
which is a candidate for the dual CFT.  Roughly,
the \nth\ twisted sector has oscillators with $O(1/n)$
moding; the energy threshold to reach this sector is $RE\sim O(n)$.
Thus the total density of states is approximately
\eqn\denst{
  \rho(E)\sim\sum_{n=0}^{\min{(k,RE)}}\rho_n
  \exp\Bigl[\beta_0\sqrt{n(RE-n)}\Bigr]
}
for some constants $\beta_0$, $\rho_n$.
For a given $E<k/2R$, the probable value of $n$ is $O(RE/2)$,
thus $\rho(E)\sim\exp[\beta_0 RE/2]$ is a Hagedorn spectrum.
We conclude that the symmetric orbifold describes 
a regime where $\lstr>\rads$ ($\gstr^2 k<1$), since
there is no regime where the system looks
like a supergravity gas in $AdS_3\times S^3$, and there is 
no 6d black hole phase.
To attain the supergravity limit requires an understanding
of the CFT away from the orbifold point.

\subsec{Falling toward a black hole}

We are now ready to consider some simple examples of objects 
falling toward a black hole
and explain how to describe them in terms of the gauge theory.%
\foot{We understand that similar ideas are being explored in \bkl.}
The basic idea is that objects initially far from the black hole are described
by localized excitations of the gauge theory. The evolution toward
the black hole is represented by an expansion of the size of the excitation.
This is a dynamical effect, and not e.g. just a change in the UV cut-off.
For definiteness, we consider the case of 
four dimensional SYM which describes string theory
in $AdS_5\times S^5$.

$AdS_5$ can be written in the form
\eqn\ads{
ds^2 = {r^2\over R^2} \left[-dt^2 + dx^2 + dy^2 + dz^2\right]
+ {R^2 dr^2\over r^2}}
where $\partial/\partial t$ becomes null at the horizon $r=0$. 
Let us compactify
$z$ with period $L$.\foot{This produces a conical singularity on the horizon,
but this difficulty will be removed shortly.}  
Any excitation of \ads\ will evolve toward $r=0$. 
For example, consider a fundamental string wound once around $z$.
If we start the string at rest at large $r$, it will collapse toward $r=0$
moving close to the speed of light.  
String theory on this background is described 
by the SYM on $S^1 \times \R^{2,1}$.  Time evolution in the
gauge theory corresponds to evolution in $t$ in \ads. 
The string at $r=\infty$ corresponds
to a Wilson loop (in the fundamental representation)
wrapping around the circle
in the gauge theory \juan. Acting on the vacuum, this Wilson
loop creates a infinitesimally thin flux tube. This state has
infinite energy, but so does  the string wound around
the infinitely long circle at $r=\infty$. A better starting point is to
introduce an UV cut-off $\delta$ and 
smear the flux tube out over a thickness
$\delta$. By the UV/IR connection (\cf\ \refs{\malda,\susswit}),
this corresponds to starting the
string at $r=R^2/\delta$. The flux tube is not a stationary state
in the SYM theory. Its width will expand in time, eventually filling all
space as $t\rightarrow \infty$. This is the analog of the string falling
toward $r=0$. Even though it takes only a finite proper time to reach
$r=0$, the coordinate $t$ diverges. This interpretation is consistent with
the above arguments that the radial direction  corresponds to a length
scale in SYM.

Suppose we now start with the near extremal black three-brane. 
The near horizon geometry is given  by
\eqn\next{
ds^2 = {r^2\over R^2} \left[-\left(1-{r_0^4\over r^4}\right) dt^2
+ dx^2 + dy^2 + dz^2\right]
+ \left(1-{r_0^4\over r^4}\right)^{-1}{R^2 dr^2\over r^2}}
It has a Hawking temperature $T= r_0/(\pi R^2)$ and an energy density
$\mu = 3\pi^2 N^2 T^4/8$.
In the gauge theory, the geometry \next\
represents a typical state with energy density $\mu$. Since the number of
quanta is large\foot{This is true for any $\mu$, since the volume in the
$(x,y)$ directions is arbitrarily large.}, such a state
is closely approximated by a thermal state with temperature $T$. 
Let us again compactify $z$, wrap a
string around it and let it fall in. In the gauge theory, 
the wrapped string again corresponds to a thin flux tube wrapped around $z$.
The flux will expand as before, but now it only expands until it reaches
the typical wavelength $1/T$ of the thermal state.  At that
point, it thermalizes. We claim that this corresponds to the wrapped
string approaching the horizon.
To see this, note that the infalling string will approximately follow a
null geodesic. So the time it takes to fall from a large
radius to $r$ of order $r_0$ is $t_0 = \int R^2 dr/r^2 = R^2/r_0$.
If we assume the flux also expands at close to the
speed of light, after this time it will have a width of order $t_0$ which is
indeed of order the thermal wavelength $t_0 = R^2/r_0 \sim 1/T$.
This agreement is quite robust: It is independent of the details of the
initial starting point, and would hold in other dimensions as well.

When the
string becomes very close to the horizon, the 
time $t$ diverges logarithmically.
It is tempting to interpret this as reflecting the approach to thermal
equilibrium in the gauge theory. Note that the ordinary Hamiltonian
evolution in the gauge theory does not  describe the string crossing the
horizon. This is not surprising, since time evolution in the gauge theory
corresponds to evolution in 
the asymptotic time $t$ in the spacetime. As we discuss below,
to see evolution across the horizon, one has to evolve the state using
an operator which is different from the usual Hamiltonian.

We have described objects falling into an existing black hole in the gauge
theory. How does
one describe the formation of the black hole itself?
A generic initial state
of high energy $E> N^2/R$
in $AdS_5$ will collapse to form a Schwarzschild-AdS black hole, which then
radiates. Since the negative cosmological constant acts like a confining box,
the black hole eventually comes into thermal equilibrium with the
gas outside. This is not an exotic process in the gauge theory on
$S^3 \times \R$.
Rather, it is simply the statement that most high
energy states will evolve into a state which is closely approximated
by the thermal state.
This is also true for states in the gauge theory with energy 
sufficiently less than $N^2/R$,
but they correspond to a gas of particles in AdS
without a black hole.

While this is the generic behavior, it is worth emphasizing that special
states can behave differently. For example, consider a low energy
supergravity excitation in AdS which is boosted 
in one direction until it has
a large energy with respect to the global time translation.
Its evolution will be approximately a geodesic which oscillates from
one side of AdS to the other. 
The point is that the choice of a time coordinate generated by
a globally timelike generator of SO(4,2) is nonunique.  A state
which is static with respect to a given Hamiltonian will be
an oscillating coherent state with respect to another.
In the SYM, this oscillating state is described by evolving
the  original SYM state 
with the conformal generator corresponding to the new global time in AdS. 
The resulting
state will no longer be static. 
Its evolution will localize it near one point
of the three-sphere, then cause it to expand and relocalize
near the antipodal point on the sphere, and then repeat. So 
changing the spatial scale is not always 
associated with a renormalization group flow: 
Sometimes the motion is reversable.


\subsec{Waves on a three-brane}

Another type of object which can be introduced into the IIB theory is
a three-brane.  One can think of the $AdS_5$ background as produced by
a large number $N$ of three-branes, leading to the representation as
large $N$ gauge theory on $\R^{3,1}$.
As discussed in \refs{\malda,\dt}, a state containing
a three-brane parallel to these at radius $r$
(in the coordinates \ads) is described by a gauge theory vacuum with
a non-zero scalar vev, having a single non-zero eigenvalue
(call it $\vec X_{11}$) with
$|\vev{\vec X_{11}}|=r$.
This breaks the gauge group to $U(N-1)\times U(1)$ and
excitations in the $U(1)$ can be interpreted as waves on the isolated brane.

Since we know the $r$ coordinate for this brane,
this identification gives us another way to produce localized
excitations in AdS.
However, to describe excitations crossing  a horizon, it would be more
interesting to have a description of the three-brane in the global
coordinates \metric, in other words as a state in gauge theory on
$S^3\times \R$.

The appropriate state is easy to find, given the fact that $\R^{3,1}$
can be conformally embedded into $S^3\times \R$.
We simply take the classical solution $\vec X_{11}=\vec x$ on $\R^{3,1}$
and apply the standard transformation law for the conformally coupled
scalar $X$.  This leads to the configuration
\eqn\xconfig{
\vev{X_{11}} = {\vec x\over \cos t + \cos \chi}
}
(where the metric on $S^3$ is
$d\Omega^2_3 = d\chi^2 + \sin^2 \chi\ d\Omega^2_2$, and $t$ is the global time).
This is a classical solution preserving $16$ of the
$32$ superconformal symmetries.  As such, it can be expected to be
an exact solution of the gauge theory.
This solution should describe the same three-brane,
but in the global coordinate system.

This solution again breaks the gauge symmetry to $U(N-1)\times U(1)$
and excitations in the $U(1)$ are again expected to be excitations of
the isolated brane.  To the extent that these excitations are confined
to the brane, they naturally propagate in Minkowski time, but
one can see how such propagation looks in global time.

More importantly, we can act on this solution with $SO(4,2)$
to obtain new three-brane
solutions.  In particular, translation of the global time variable
produces the solutions
\eqn\xconfigtwo{
\vev{X_{11}} = {\vec x\over \cos (t-t_0) + \cos \chi}
}
which extend out of the region covered by the coordinates \ads.
(A stationary three-brane in the coordinates \ads\ will asymptote to the
boundary of this region at large $r$.
Since the boundary is at $\cos t = 1/\sqrt{1+r^2}$, any shift of $t$
will cause the brane to cross the boundary.)

This is an example of a brane solution which crosses the horizon of an
extreme black hole.  Is it possible to construct a brane which crosses
the event horizon of a nonextreme black hole background?

\subsec{Falling into BTZ black holes}

We have seen  in sec. $4.3$
that when an AdS black hole is represented as a
state in the 
dual CFT, the natural Hamiltonian evolution corresponds to evolving with
respect to the external Schwarzschild time of the black hole. On the other
hand, since the global time in AdS is not unique, it is natural to
 consider different evolutions in the spacetime which correspond to 
 different operators in the CFT playing the role of the Hamiltonian.
 Combining these ideas, one is led to consider alternative evolutions in
 the black hole case.
 In particular, one might ask whether there is an
operator which approximates the experience of an 
infalling observer. We will argue below that the answer is yes. The fact
that there is a single Hilbert space, with different (noncommuting)
operators  describing external and infalling observers, is 
a concrete realization of the ideas of black hole complementarity \stu.

An ideal context to illustrate this
is the BTZ black hole \btzbh, whose spacetime geometry
is locally $AdS_3$. Some of the local isometries have orbits which 
cross the horizon, so the corresponding conformal generator
in the CFT is a natural
candidate for evolving the state for these infalling observers.
To be specific, we start with the description of 
$AdS_3$ as the hyperboloid
\eqn\adss{-T_1^2 - T_2^2 + X_1^2 + X_2^2 = -R^2}
in $\R^{2,2}$. A convenient way to parameterize this surface is to
introduce coordinates associated with two commuting symmetries. If we
let $t,\pp$ parameterize rotations in the $T_1,T_2$ and $X_1,X_2$
planes respectively, then the metric takes the standard form
\eqn\aads{ ds^2 = -\left({r^2\over R^2} +1\right) dt^2 +
\left({r^2\over R^2} +1\right)^{-1} dr^2 + r^2 d\pp^2}
If, instead, we let $t,\pp$ parameterize boosts in the $T_1,X_1$ and $T_2,X_2$
planes respectively, the metric takes the form
\eqn\abtz{ ds^2 = -\left({r^2\over R^2} -1\right) dt^2 +
\left({r^2\over R^2} -1\right)^{-1} dr^2 + r^2 d\pp^2}
The BTZ black hole is obtained by periodically identifying $\pp$ in \abtz,
and the total energy depends on the choice of period.\foot{We 
restrict our attention to nonrotating black holes;
there is a simple generalization to the rotating case
involving periodically identifying a linear combination of $t$ and $\pp$.}
Alternatively, one can rescale $r$
so that the period of $\pp$ is always $2\pi$ and the metric becomes:
\eqn\gbtz{ ds^2 = -\left({r^2\over R^2} -m\right) dt^2 +
\left({r^2\over R^2} -m\right)^{-1} dr^2 + r^2 d\pp^2}
The total energy of this solution relative to the $AdS_3$ ground state
\aads\ is related to the parameter $m$ by $E\sim E_{BTZ} + mk/R$, where
$k=Q_1 Q_5$ and $E_{BTZ}= k/R$ as in \hithree. The Hawking temperature is
$T=\sqrt m/(2\pi R)$. Note that the form of the metric \gbtz\ also includes
$AdS_3$ (without identifications) by setting $m=-1$.

We have seen that a rotation in the $T_1,T_2$ plane corresponds to 
the global time translation in $AdS_3$ \aads, but not in the BTZ black hole
\gbtz\ (with $m>0$). Nevertheless, in the latter case,
it is still a local isometry which is
simply not invariant under translations of $\pp$ by $2\pi$. When $m>0$, the 
radial coordinate in \gbtz\ is related to  the embedding variables by
$r^2 = m(T_2^2 - X_2^2)$. So if we start with $r^2>0$ and
$T_1=0$ (which corresponds to 
$t=0$ in \gbtz) and rotate $T_1,T_2$ keeping $X_1,X_2$ fixed, then
$r$ decreases to zero after a finite rotation. In other words,  the orbits of
the local symmetry $T_1 \d/\d T_2 - T_2 \d/\d T_1$ are timelike curves which
fall into the black
hole and hit the singularity in finite proper time. 

How is this described in the CFT? The usual Hamiltonian, $L_0 + \bar L_0$,
generating time
translations on $S^1 \times \R$, always corresponds to
translations of $t$ in \gbtz. This is true for either sign of $m$. Similarly,
$L_0 - \bar L_0$ always corresponds to translations of $\pp$. Thus
even though the local isometries of the spacetime,
$SO(2,2)$, agree with the conformal symmetries of the $1+1$ dimensional
CFT, the relation between the generators depends on the energy.
For $m<0$ ($E<E_{BTZ}$), the spacetime generator $T_1 \d/\d T_2 - T_2 \d/\d T_1$
is represented in the CFT by $L_0 + \bar L_0$,  but for $m>0$ ($E>E_{BTZ}$),
it is not.  Since
$L_0 + \bar L_0$ and $L_0 - \bar L_0$  now correspond to boosts in the
$T_1,X_1$ and $T_2,X_2$ planes, the algebra implies that 
\eqn\infallgen{
  T_1 {\d\over \d T_2} - T_2 {\d\over \d T_1} = {1\over 2}
	(L_1+L_{-1}+{\bar L}_1+{\bar L}_{-1})\ .
}

If we adjust the constants in $\lz,\lzb$ so that they vanish for the NS ground
state,
the BTZ black hole is represented on the CFT side
by the ensemble of states with energy 
$\lz=\lzb>k/4$.  In this regime, the canonical and
microcanonical ensembles are equivalent.
In $\AdS5s5$, the transition from a gas of particles
in AdS to a black hole in AdS is associated with a 
`deconfinement' transition in the gauge theory \wittenads.
In $AdS_3$, the analogous order parameter involves the 
permutation symmetry of the symmetric product
$S^{k}(M)$ \maldatalk; singlets dominate the gas phase,
while large representations dominate the BTZ phase.
At finite $k$, this crossover is not singular,
and the CFT `sums' over all topologies in supergravity that
are asymptotically $AdS_3\times S^3\times M$
(including 6d Schwarzschild black holes, as discussed above).
At large $k$, and well into the BTZ regime $\lz,\lzb\gg k/4$,
the classical BTZ spacetime is the only relevant geometry.
As we have said, the time conjugate to $\lz+\lzb$ is to be related
to the static external time of the black hole solution.
In direct analogy with the $\AdS5s5$ case, a probe dropped into the black hole
will be seen to be thermalized in the CFT as it evolves
according to the asymptotic time variable.  However,
here the bulk spacetime is locally $AdS_3$, so we can
also evolve the CFT state using \infallgen, corresponding
to an infalling frame of reference.  The black hole state
in the CFT is not an eigenstate of this operator,
corresponding to the fact that the black hole geometry
is not static in the infall frame. 
The fact that the operator \infallgen\ is not invariant under $2\pi$ shifts
of $\pp$ in the BTZ spacetime is not a problem, since all of the generators
of the conformal symmetry act on all of the states in the CFT -- even those
which are not invariant.
To summarize, the evolution operators for
infalling and static observers simply
correspond to different (noncommuting) generators of SO(2,2),
whose action on the CFT states is canonical.

Neither evolution is singular in the finite $k$ CFT.
The static evolution is just that -- the black hole states
are eigenstates of $\lz+\lzb$.  The infall evolution is
also nonsingular, since any SO(2,2) generator
acts unitarily on the CFT Hilbert space.
This shows that, {\it given the AdS/CFT duality, quantum
effects indeed resolve the BTZ black hole singularity!}
The infall evolution generator \infallgen\
should develop a singularity in the classical
limit $k\rightarrow\infty$, $E>k/4$, in the regime
$\lstr\ll\rads$ where one can trust supergravity; this possibility is
currently under investigation.  Such a singularity 
would simply reflect an impropriety of the limit (as in
geometric optics near a focal point)
rather than of the theory itself.
The `long string' model
that pertains to the orbifold locus in the CFT moduli
space does not appear to exhibit such a singularity
for finite infall evolution parameter even in this limit,
perhaps because this CFT describes \adsthree\ with
$\lstr>\rads$, so that the geometry is stringy.
Details will appear in \bdhmii.

In higher dimensions, we believe that a similar story should hold:
Observers that fall into a black hole are described
in the CFT by evolving the state by an operator (or family of operators)
which are different from the usual Hamiltonian. To determine this operator
and find the behavior of a black hole state near the curvature 
singularity is an outstanding problem.
In this regard, it is interesting to note the following.
The above observation that the 
relation between the spacetime and CFT symmetry generators depends
on the energy of the state, 
seems to  be a special feature of $2+1$ dimensions.
In higher dimensions, 
the Schwarzschild-AdS black hole does not have constant curvature.
But in the asymptotically AdS region, 
the time translation symmetry outside the black hole
is the same as a global time translation in AdS, \ie, a compact generator
of $SO(n,2)$. This can be seen from the fact that the boost symmetries do 
not commute with all
the rotation symmetries $SO(n)$ present outside the black hole.

A different approach to some of the questions addressed in this paper has
recently been given in \berk.

\vskip 1cm
\noindent{\bf Acknowledgements:}
We thank the participants of the ITP workshop
`Dualities in String Theory' for discussions, and the
organizers for generating the stimulating environment 
in which the bulk of this work was carried out.  T.B. and E.M. similarly
thank the organizers and participants of the Amsterdam Summer
Workshop, `String Theory and Black Holes', and
in particular, J. Maldacena. 
Preliminary reports on our results were presented at Strings'98
(http://www.itp.ucsb.edu/online/strings98/banks),
and at the Spinoza meeting on the Quantum Black Hole,
Utrecht, June 30-July 4.

This work was supported in part by
NSF grants PHY94-07194 and PHY95-07065,
and DOE grants DE-FG02-96ER40559 and DE-FG02-90ER-40560.


\listrefs
\end


\ref\opcl{E. Cremmer and J. Scherk, \npb{50}{1972}{222}..}; 
\ref\induced{E. Witten, \npb{276}{1986}{291}.} 
\ref\bfss{T. Banks, W. Fischler, S.H. Shenker and L. Susskind,
hep-th/9610043; \prd {55}{1997}{5112}.}, 

\nref\gkp{S.S. Gubser, I.R. Klebanov, and A.M. Polyakov, hep-th/9802109.}%
\refs{\malda,\gkp,\wittenads}
\nref\dodf{M.R. Douglas, hep-th/9604198;
R. Dijkgraaf, E. Verlinde, and H. Verlinde,
hep-th/9704018, \npb{506}{1997}{121};
E. Witten, hep-th/9707093;
S.F. Hassan and S. Wadia, hep-th/9703163,
\plb{402}{1997}{43}; hep-th/9712213.}%
\nref\maldathesis{J. Maldacena, Princeton Ph.D. thesis, hep-th/9607235;
see also hep-th/9705078, Nucl.Phys.Proc.Suppl. {\bf 61A} (1998) 111-123.}%

\nref\hipt{A. Achucarro and P.K. Townsend, \plb{180}{1986}{89};
P.S. Howe, J.M. Izquierdo, G. Papadopoulos, and
P.K. Townsend, hep-th/9505032.}%
\nref\witcsgrav{E. Witten, \npb{311}{1988}{46}; \npb{323}{1989}{113}.}%
to the corresponding WZW theory on the boundary
\ref\wzw{E. Witten, \cmp{121}{1989}{351};
G. Moore and N. Seiberg, \plb{220}{1989}{422};
S. Elitzur, G. Moore, A. Schwimmer, and N. Seiberg, \npb{326}{1989}{108};
W. Ogura, \plb{229}{1989}{61}.};
\nref\polgrav{A.M. Polyakov, \mpl{A2}{1987}{893};
V.G. Knizhnik, A.M. Polyakov, A.B. Zamolodchikov \mpl{A3}{1988}{819}.}%
\nref\verlinde{H. Verlinde, \npb{337}{1990}{652}.}%
\nref\carliou{S. Carlip, \npb{362}{1991}{111}.}%

\nref\dasmathur{S.R. Das and S.D. Mathur, hep-th/9606185,
\npb{478}{1996}{561};
A. Dhar, G. Mandal, and S. Wadia, hep-th/9605234,
\plb{388}{1996}{51}.}%
\nref\maldastrom{J. Maldacena and A. Strominger, hep-th/9609026,
\prd{55}{1997}{861}.}%
\nref\maldastromang{J. Maldacena and A. Strominger, hep-th/9702015;
\prd{56}{1997}{4975}.}%
\nref\mathurang{S.D. Mathur, hep-th/9704156.}%
\nref\gubser{S. Gubser, hep-th/9704195; \prd{56}{1997}{4984}.}%
\nref\fxsclrs{C.G. Callan, S.S. Gubser, I.R. Klebanov, and A.A. Tseytlin,
hep-th/9610172, \npb{489}{1997}{65}.;
I.R. Klebanov and M. Krasnitz, hep-th/9612051, \prd{55}{1997}{3250};
I.R. Klebanov and M. Krasnitz, hep-th/9703216, \prd{56}{1997}{2173};
I.R. Klebanov, A. Rajaraman, and A.A. Tseytlin, hep-th/9704112,
\npb{503}{1997}{157}.}%
\ref\noncrit{P. Ginsparg and G. Moore, {\it Lectures on
2d gravity and 2d string theory}, hep-th/9304011; in Recent
Directions in Particle Theory (TASI 1992), J. Harvey and J. Polchinski
(eds.), World Scientific (1993).});

\ref\stromvafa{A. Strominger and C. Vafa, hep-th/9601029;
\plb{379}{1996}{99}.}.
\ref\spinbh{J.C. Breckenridge, D.A. Lowe, R.C. Myers, 
A.W. Peet, A. Strominger, C. Vafa, hep-th/9603078,
\plb{381}{1996}{423};
J.C. Breckenridge, R.C. Myers, A.W. Peet, C. Vafa, hep-th/9602065,
\plb{391}{1997}{93}.}.
\ref\btz{M. Ba\~nados, C. Teitelboim, and J. Zanelli,
\prl{69}{1992}{1849}.  For an extensive review and further
references, see S. Carlip, gr-qc/9506079;
\cqg{12}{1995}{2853}.}.  
\ref\garf{D. Garfinkle and T. Vachaspati, \prd{42}{1990}{1960};
D. Garfinkle, \prd{46}{1992}{4286}.}.
\nref\bh{J.. Brown and M. Henneaux, \cmp {104}{1986}{207}.}%
\nref\banados{M. Ba\~nados, hep-th/9405171; \prd {52}{1996}{5816}.}%
\nref\strom{A. Strominger, hep-th/9712251.}%
\nref\bbo{M. Ba\~nados, T. Brotz, and M. Ortiz, hep-th/9802076.}%
\ref\chvd{O. Coussaert, M. Henneaux, and P. van Driel, 
gr-qc/9506019; \cqg {12}{1995}{2961}.}
\ref\aleks{A. Alekseev and S. Shatashvili, \npb{323}{1989}{719};
M. Bershadsky and H. Ooguri, \cmp{126}{1989}{49}.}.
\ref\kpr{C. Kounnas, M. Porrati, and B. Rostand,
\plb{258}{1991}{61}.}.  
\ref\cousshen{O. Coussaert and M. Henneaux, hep-th/9310194; 
\prl{72}{1994}{183}.};
\ref\bssent{D. Birmingham, I. Sachs, and S. Sen, hep-th/9801019.}%
\nref\otherent{V. Balasubramanian and F. Larsen, hep-th/9802198;
N. Kaloper, hep-th/9804062.}%
\ref\kutseib{D. Kutasov and N. Seiberg, \npb{358}{1991}{600}.},
\ref\carlip{S. Carlip, gr-qc/9409052; \prd{51}{1995}{632}.},
\ref\hormarolf{G. Horowitz and D. Marolf, hep-th/9605224; 
\prd{55}{1997}{835}.}
\ref\bis{D. Birmingham, I. Sachs, and S. Sen, hep-th/9707188.}.
\ref\thooft{G. 't Hooft, \npb{256}{1985}{727}.}).
\ref\mss{G. Moore, N. Seiberg, and M. Staudacher, \npb{362}{1991}{665}.}.
\ref\cghs{C. Callan, S. Giddings, J. Harvey, A. Strominger,
hep-th/9111056; \prd{45}{1992}{1005}.}.
\ref\mirror{T.-D. Chung and H. Verlinde, hep-th/9311007; 
\npb{418}{1994}{305}.}.
ref\limart{M. Li and E. Martinec,
hep-th/9703211, \cqg{14}{1997}{3187};
hep-th/9704134, \cqg{14}{1997}{3205};
hep-th/9709114.}
\ref\gubkleb{S.S. Gubser and I.R. Klebanov, hep-th/9708005.}
(see also
\ref\fmmr{D.Z. Freedman, S.D. Mathur, A. Matusis, and L. Rastelli,
hep-th/9804058.}).